\documentclass[showpacs,preprintnumbers,prl,a4paper,twocolumn]{revtex4}
\usepackage{amssymb}
\usepackage{amsfonts}
\usepackage{amsmath}
\usepackage{graphicx}
\usepackage{dcolumn}
\usepackage{bm}

\begin{document}

\preprint{}
\title{\Large\bf Mapping from Architecture to Dynamics: A Unified View of Dynamical Processes on Networks}
\author{Jie Zhang $^1$ }
\email[]{enzhangjie@eie.polyu.edu.hk}
\author{Changsong Zhou $^2$}
\author{Xiaoke Xu $^3$}

\author{Michael Small $^1$}

\affiliation{$^1$Department of Electronic and Information
Engineering, Hong Kong Polytechnic University, Hong Kong, P. R.
China
 $^2$ Department of Physics, Center for nonlinear
studies, Hong Kong Baptist University, Hong Kong, P. R. China
 $^3$
School of Communication and Electronic Engineering, Qingdao
Technological University, Qingdao, P. R. China}
\begin{abstract}
Although it is unambiguously agreed that structure plays a
fundamental role in shaping the dynamics of complex systems, this
intricate relationship still remains unclear. We investigate a
general computational transformation by which we can map the network
topology directly to the dynamical patterns emergent on it ---
independent of the nature of the dynamical process. We find that
many seemingly diverse dynamical processes such as coupled
oscillators and diffusion phenomena can all be understood and
unified through this same procedure. Using the multiscale complexity
measure derived form the structure-dynamics transformation, we find
that the topological features like hierarchy, heterogeneity and
modularity all result in higher complexity. This result suggests a
universal principle: it is the desire for functional diversity that
drives the evolution of network architecture.

\end{abstract}
 \maketitle

Recent advances in the realm of complex networks have furnished us
with a new paradigm to understand and characterize complex systems
\cite{BA1,Newman,Boccaletti}, from the human brain that is composed
of billions of interconnected neurons, to our society with six
billion cooperating individuals. The discovery of the distinctive
scale free \cite{scalefree} and the small world \cite{smallworld}
structures has fundamentally altered our view of these networks. A
variety of other complex topological features, such as high
clustering \cite{Q}, hierarchical ordering \cite{hiera}, and degree
mixing \cite{assort1}, are also emerging as important to the overall
behaviour of network systems. However, recent progress has focussed
mainly on the underlying topology \cite{BA1}, the effort to unravel
the system's dynamics or function has been less advanced
\cite{dynamical process,Newman,Boccaletti}. With the increasing
capability to capture simultaneously the time dependent activity of
many components in complex systems \cite{netdyna} (such as the
multiple electrode recordings and gene expression patterns),
unraveling the intricate relationship between the structural and
functional characteristics has become a problem of utmost
importance, and hints toward generic organizing principles and a
deeper understanding of ``complexity". Generally, the topological
descriptors fail to explicitly capture the dynamical aspects. In
order to characterize the dynamics, one must implement the
particular dynamical process on networks via extensive numerical
simulations.

In this paper we propose a methodology that allows one to predict
the collective dynamics (or functions) of a complex system directly
from the underlying topology. Yet, this methodology is
\emph{independent  of the details of dynamical process}. This is
achieved by constructing \emph{node interaction profiles} through a
kernel function, which quantitatively identify the role of each node
in shaping the integrated dynamics and thus captures the way nodes
are dynamically interacting with one another. Notably, we find that
synchronization of coupled oscillator, ensemble neuron firing,
epidemic spreading and diffusion process can all be unified under
this theoretical picture, suggesting that some universal mechanisms
may govern the overall dynamical behavior of the seemingly diverse
complex systems. Based on this transformation we propose a
``function-driven" multiscale complexity measure by virtue of the
adjustable kernel bandwidth, which unravels the functional
organization of a network at different levels. We find that
structural complexity measures such as topological heterogeneity
\cite{BA1},modularity \cite{Q}, hierarchy \cite{hiera}, and
nontrivial correlation \cite{assort1} all translate into higher
functional complexity, indicating that the need for multiple
function governs the structural evolution of networks.




Consider a complex system of $N$ coupled dynamical units, whose
equations are described by \cite{carroll}:
$\dot{x_i}=F(x_i)+\sigma\sum_{j=1}^N A_{ij}H(x_j),    i=1,2,...,N$
where $\dot{x_i}=F(x_i)$ governs the dynamics of each component, $H$
is a fixed output function, $\sigma$ represents the coupling
strength, and $A$ is the coupling or adjacency matrix. The problem
is now, given the topology $A$ of the network, can we infer
qualitatively the dynamics without implementing and computationally
simulating it? If so, what does this tell us about the behavior of
different dynamical units within the same network structure? By
``dynamics" we refer to the pattern of temporal correlations among
outputs $x_{i}(t), i=1,...,N$. This has been intensively studied in
chaos community and in brain research groups, under either the
banner of \emph{synchronization} \cite{carroll,zongshu,Arenas1} or
\emph{functional connectivity} \cite{Spornss}.

Obviously, to predict the collective dynamics, we must identify the
effective influence of each unit to others, and distinguish their
specific, functional roles in shaping the dynamics of each unit. A
fundamental feature of complex interacting systems is the presence
of interactions across all scales. To describe the fact that
directly coupled components usually exert a stronger influence on
each other and the impact attenuation through intermediaries nodes,
we adopt a simple, monotonically decreasing function known as a
``kernel" $K$. For a pair of nodes $i$ and $j$, we then define their
effective interaction $R_{ij}$, in the form of
$K\left(x_{ij},{h}\right)$, where $x_{ij}$ is their shortest
distance, $K$ is a non-negative, symmetric kernel function
satisfying $\int_R K(x)d(x)=1$, $\int_R x K(x)d(x)=0$,
$\lim_{x\rightarrow \infty} K(x)d(x)=0$, and $h$ is the
\emph{bandwidth} that controls the width of the kernel. For Gaussian
kernel, the interaction matrix $R$ will read: $R_{ij}=\exp
(-{x_{ij}^2}/{2h^2})$.

Note that the $i$-th row of  $R$, $R_{i}$ portrays the
\emph{effective interaction} node $i$ receives from all its
neighbors, which we call \emph{interaction profile} of node $i$.
This vectorial profile systematically encodes the identity of all
neighbors of node $i$ as distinct driving forces (with different
intensities determined by the kernel) to its own dynamics, and does
not depend on specific choice of kernels. Therefore $R_{i}$ defines
the unique ``status" of node $i$. To further predict the dynamical
correlation or functional connectivity $F_{ij}$ between node $i$ and
$j$, we can calculate the similarity between their interaction
profiles $R_i$ and $R_j$: $ F_{{ij}}= {{R_{i}\cdot R_{j}}\over
{|R_{i}| |R_{j}}|} $

The similarity $F_{ij}$ provides a unique clue to evaluate the
dynamical proximity between the components. Unit $i$ and $j$ subject
to a large number of common inputs (up to higher orders) are more
likely to behave similarly. In this case, their profiles $R_{i}$ and
$R_{j}$ will largely coincide by sharing many common entries,
leading to a large $F_{ij}$
--- approaching $1$. Conversely, a pair of units with few common
drives tend to be independent and thus have a $F_{ij}$ near 0. A
great advantage of this ``kernel" formalism lies in the adjustable
bandwidth $h$ which can evaluate different levels of function of the
network at various topological scales. In the following we will
demonstrate how the collective dynamics of various dynamical
processes can be predicted using the above mapping from structure
$A$ to dynamics (or function) $F$.

\begin{figure}[htbp]
\begin{center}
\includegraphics[height=0.25\textwidth]{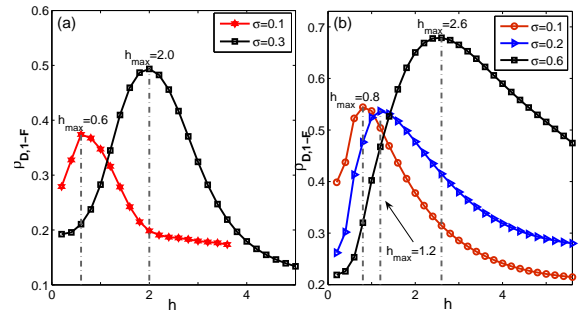}
\end{center}
\caption{The correlation coefficient $\rho$ between $D_{ij}$ and
$1-F_{ij}$ for coupled phase oscillator on (a) random (ER) network
with 500 nodes (mean degree is 5) and (b) a collaboration network
with 379 nodes, where nodes are scientists who conduct research on
networks and links represent coauthorship \cite{netsci}.}
\label{netsci}
\end{figure}

\begin{figure}[htbp]
\begin{center}
\includegraphics[height=0.2\textwidth]{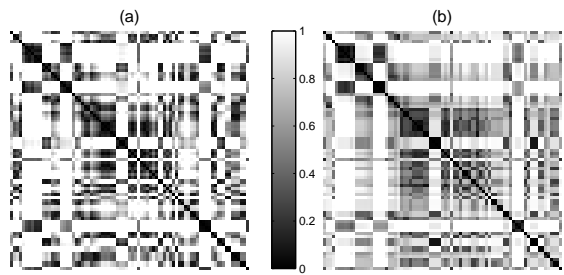}
\end{center}
\caption{Visualization of matrixes (a) $D_{ij}$ and (b) $1-F_{ij}$
obtained form Fig. \ref{netsci} (b), i.e., phase oscillators on
collaboration network, with $\sigma=0.6$ in computing $D_{ij}$ and
$h=2.6$ for $F_{ij}$.} \label{adj2}
\end{figure}

We start with synchronization phenomena, which are widely observed
in nature and occupy a privileged position in understanding
collective behavior in various disciplines \cite{zongshu}. Recently
the interplay between a network's structure and its synchronization
dynamics has attracted significant attention \cite{Arenas1,Kahng}.
Here we use Kuramoto model defined on various networks as a
prototype example. It is govern by $\dot{\theta_{i}}=\omega_{i}+
{\sigma \over {\bar{k}}}A_ {ij} \sum_{j=1} ^{N}
\sin(\theta_j-\theta_i), i=1,2,..., N$, where $\omega_{i}$ is the
frequency of phase oscillators (uniformly distributed in (0,1)),
$\sigma$ is the coupling strength, $A$ is the adjacency matrix, and
$\bar{k}$ is the mean degree. Specifically, we will show that
$F_{ij}$ obtained at different kernel bandwidth $h$ can provide a
good prediction of how the collective dynamics evolves with
$\sigma$.

With a small coupling $\sigma$, the oscillators are mostly
independent. The dynamical distance between outputs of node $i$ and
$j$, defined as $ D_{ij}=\langle\theta_i(t)-\theta_j(t)\rangle $
($\theta(t)$ are wrapped to $[0,2\pi]$ and $\langle\cdot\rangle$
means time average) will be non-zero and constitute a narrow
distribution. When $\sigma$ is large, the whole network reaches
complete synchronization, and $D_{ij}$ distribution will be a narrow
peak again near $0$. For intermediate $\sigma$, various functional
clusters are formed, with $D_{ij}$ distribution broadening. We find
that the $D_{ij}$ distributions at various $\sigma$ are exactly
reproduced by $F_{ij}$ using different $h$. At a small $h$, all
entries in $R_i$ are almost $0$ except the $i$th, meaning each node
only has impact on itself. The $R_i$s are mostly orthogonal, thus
$F_{ij}$ will centralize at $0$. By contrast, the kernel becomes
flat at a large $h$, leading every node to exert similar influence
on all others. The $F_{ij}$ then concentrates near $1$ as all $R_i$s
are almost identical. For medium $h$, the distribution of $F_{ij}$
broadens within $[0,1]$ with the peaks corresponding to the formed
functional clusters.

These examples show that the kernel bandwidth $h$ is playing a role
directly analogous to coupling strength $\sigma$, and $F_{ij}$
offers a good prediction of the collective dynamics $D_{ij}$. To
further verify this, we first get $D_{ij}$ by implement phase
oscillators on networks (a random network and a modular
collaboration network \cite{netsci} are used here as examples). We
then plot the correlation coefficient $\rho$ between $D_{ij}$ and
$1-F_{ij}$ that is obtained at various $h$, see Fig. \ref{netsci}
(we use $1-F_{ij}$ because it measures dissimilarity similar to
$D_{ij}$, while $F_{ij}$ is a similarity measure). We find that for
a given $\sigma$, there is always an optimal kernel bandwidth
$h_{max}$ that attains a maximum similarity between $D_{ij}$ and
$1-F_{ij}$, with $h_{max}$ being proportional to $\sigma$. The
matrices $D_{ij}$ and $1-F_{ij}$ demonstrate vary similar patterns
(see Fig. \ref{adj2}), reflected by a large correlation coefficient
$\rho$. We find that $\rho$ generally takes a large value for medium
and high coupling $\sigma$, where oscillators have self-organized
into functional clusters. For weak $\sigma$, the oscillators are
largely independent, thus $D_{ij}$ is somewhat random and cannot be
accurately fitted by $F_{ij}$.

Now we turn to a concrete example in neuroscience, the coupled
neural oscillators in cortex, which communicate by non-smooth,
pulse-like firings. The population dynamics of the neurons, like the
synchronous firing plays a vital role in cognitive function of the
brain. Therefore understanding how connectivity patterns influence
the emergent dynamics is of special concern in neuroscience. Here we
try to approach the population neuronal dynamics directly from the
underlying anatomy. In particular, we couple the FitzHugh- Nagumo
neurons through real networks (by excitatory synapse with synaptic
conductance $g$) to get $D_{ij}$ and check if it can be predicted by
$F_{ij}$. We use two networks possessing key topological properties
of the cortex, i.e., small-world, hierarchical and modular
structure. One is the neural network of \emph{Caenorhabditis
elegans} whose anatomy has been identified. The other is a
hierarchically organized modular network \cite{Arenas1} with two
hierarchical levels.


\begin{figure}[htbp]
\begin{center}
\includegraphics[height=0.25\textwidth]{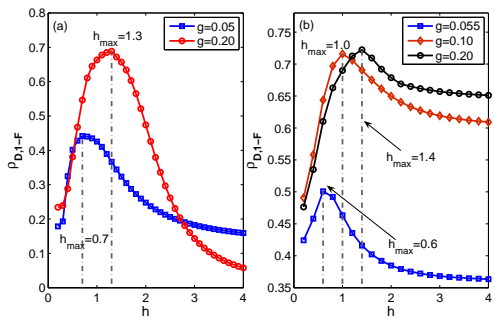}
\end{center}
\caption{The correlation coefficient between $D_{ij}$ and $1-F_{ij}$
for (a) C. elegans network with 297 neurons and each has 14 synaptic
couplings on average. (b) Hierarchical network with 480 nodes
\cite{Arenas1}. Each node has 20 links to the most internal
community (formed by 30 nodes), 2 links to the most external
community (120 nodes that form four 30-nodes groups), and 1 more
link to any other node.  See supplement material for the detailed
parameters of the FHN neurons.} \label{fhn_quan}
\end{figure}

The synaptic conductance $g$ determines the amplitude of pulse
conducted to post-synaptic neurons. It plays the same role as
$\sigma$ in coupled phase oscillators. The neurons fire almost
randomly with a small $g$, and begin to form synchronous firing as
$g$ increases, giving rise to coherent oscillations. We define the
collective dynamics of neuron firing as
$D_{ij}=\langle\|f_i(n)-f_j(n)\|\rangle$, where $f_i(n)$ is the
number of firings within time window $n$ for neuron $i$. We find
that the functional similarity $F_{ij}$ obtained purely from the
network structure shows a high correlation with $D_{ij}$ for both
the \emph{C. elegans} and the hierarchical network, see Fig.
\ref{fhn_quan}. Therefore we can precisely predict the dynamics of
neuronal populations. Moreover, the kernel bandwidth $h$ plays a
role similar to synaptic conductance $g$. For a given $g$, there is
always an optimal bandwidth $h_{max}$ under which $F_{ij}$ best fits
$D_{ij}$, and this $h_{max}$ is proportional to $g$.

\begin{figure}[htbp]
\begin{center}

\end{center}
\includegraphics[height=0.25\textwidth]{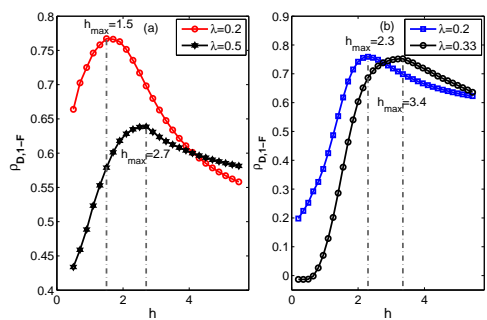}
\caption{ Correlation coefficient between $D_{ij}$ and $1-F_{ij}$
for epidemic spreading on (a) scale free BA network (500 nodes with
mean degree 5) and (b) email network \cite{emailnet} (1133 nodes).}
\label{epi_quan}
\end{figure}

The above mapping can predict the collective behavior not only for
the coupled dynamical systems, but also for the general diffusion
process on networks like epidemic spreading \cite{epi}. The
collective behavior here means the correlation among the epidemic
dynamics of the individuals, with the dynamics of node $i$ being a
discrete-time stochastic process $S_i(t)$ ($1$ indicates infected
and $0$ for healthy). The collective dynamics is then defined as
$D_{ij}=\langle\|S'_{i}(n)-S'_{j}(n')\|\rangle$, where $S'_i(n)$ is
coarse grained from $S_i(t)$ by counting the number of $1$ in time
window $n$. We find that $D_{ij}$ is again nicely predicted by
$F_{ij}$ under a suitable $h$, see Fig. \ref{epi_quan}, where we run
SIS models on two typical networks. Interestingly, we find $h_{max}$
relates closely to the effective spreading rate $\lambda=\nu /
\delta$, where $\nu$ and $\delta$ are infection and recovery rate of
an individual. In Fig. \ref{epi_quan} we see that $D_{ij}$ under
small and large $\lambda$ is best fitted by a kernel with small and
large $h$, respectively. This is because an infected node recovers
quickly at a small $\lambda$ and has a small influence range. A
large $\lambda$ makes the node persistently infective to even its
higher order neighbors, thus is described better by a wider kernel.

Having established the mapping from structure to function, we are
naturally led to the fundamental problems: why do the various
structures such as modularity, hierarchy and degree mixing exist in
real networks? What are their roles in shaping the function? Here we
provide an explanation under the framework of complexity by
exploiting the above structure-function transformation.
Understanding complexity \cite{tame,complexity} has long been a
grand challenge that spans a wide variety of fields. Sporns et. al.
proposed ``neural complexity" to measure functional connectivity by
implementing Gaussian dynamics \cite{Sporns} on networks.
Evolutionarily speaking, the survival of a complex system hinges
crucially on the versatility of its function. Inspired by this, we
propose to evaluate network complexity by the dynamical patterns it
can support. Recall that in coupled oscillators, the evolution of
the collective dynamics versus coupling strength are well captured
by $F_{ij}$ under various $h$, thus we can approach the function
directly through $F_{ij}$. Here we use entropy, defined as $H(F) =
\sum_i^m -p_i\log(p_i)$, to characterize the diversity of $F_{ij}$
obtained at different $h$. We call it \emph{multiscale entropy}, as
it portrays the functional complexity of the network at various
topological scales.


\begin{figure}[htbp]
\begin{center}
\includegraphics[height=0.23\textwidth]{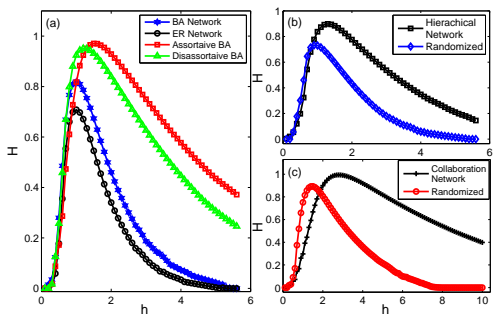}
\end{center}
\caption{Multiscale entropy for (a) scale free BA, random ER
network, assortively and disassortatively mixed BA network. All
networks have $1000$ nodes and mean degree $20$. (b) Hierarchical
networks (used in Fig. \ref{fhn_quan}) and (c) Collaboration network
\cite{netsci}. Here we normalize $H$ by a factor $H_m=log(m)$, which
is the entropy for uniform distribution, and $m$ is the bin number.}
\label{entropy}
\end{figure}

Now we examine how the intricate topologies can influence the
network complexity by computing multiscale entropy. As can be seen
in Fig. \ref{entropy}(a), the degree heterogeneity (scale free
distribution) and degree mixing \cite{assort1} (assortive and
disassortative) both lead to higher $H$ at various scales, as they
promote differentiation or facilitate the formation of modules that
can engage into various functions. Remarkably, we find that
hierarchical modular networks, which are widely observed in
biological and social systems \cite{hiera}, demonstrate notably
higher complexity than their random counterparts, see Fig.
\ref{entropy}(b) and (c). This is because the multiscale modular
structure can provide different levels of function persistently at
different scales. The fact that all these distinctive topologies
lead to higher functional complexity provides important insights
into the evolutionary mechanism of real networks. This suggests a
common principle: the demand for functional capability, shapes the
network architecture during the development of a physical network.







In summary we have introduced a novel approach that can map the
topological structure of a network directly to its functional
organization independent of the details of dynamical processes. Our
method not only provides a good prediction of the collective network
dynamics, but also allows us to conceptualize the ``complexity" of
networks through the emergent functions conveniently. This mapping
can furthermore be considered as a promising scheme to other
problems like inverse engineering and controllability. The explicit
relation between structure and function will provide unique clues to
infer the structure back from dynamical patterns, possibly with
constraints like sparseness of connectivity. The control over the
general networked systems can also be expected to enhance
conveniently by locating the most sensitive nodes or links, the
removal or rewiring of which would result in better network
performance.

\begin{acknowledgments}
This work is funded by the Hong Kong Polytechnic University
Postdoctoral Fellowships Scheme 2007-08 (G-YX0N). CS Zhou is
supported by Hong Kong Baptist University.
\end{acknowledgments}

\end{document}